\documentclass[3p,preprint]{elsarticle}

\usepackage{amssymb}

\usepackage{amsmath} 

\usepackage{amssymb,amsthm,MnSymbol}
\usepackage{color}

\usepackage{algorithm}
\usepackage{algorithmic}

\usepackage{subfig}

\usepackage{amscd}
\begin{document}
\begin{frontmatter}

\author[addr1]{I. Shevchenko\corref{cor1}}
\ead{i.shevchenko@imperial.ac.uk}
\author[addr1,addr2]{P. Berloff}
 
\cortext[cor1]{Corresponding author at:}
\address[addr1]{Department of Mathematics, Imperial College London, Huxley Building, 180 Queen's Gate, London, SW7 2AZ, UK}
\address[addr2]{Institute of Numerical Mathematics of the Russian Academy of Sciences, Moscow, Russia}


\title{A method for preserving nominally-resolved flow patterns in low-resolution ocean simulations: Constrained dynamics}




\begin{abstract}
Inability of low-resolution ocean models to simulate many important aspects of the large-scale general circulation is a common problem.
In the view of physics, the main reason for this failure are the missed dynamical effects of the unresolved small scales of motion on the explicitly resolved large scale motions.
Complimentary to this mainstream physics-based perspective, we propose to address this failure from the dynamical systems point of view, namely, as the persistent tendency of phase space trajectories representing the low-resolution solution to escape the right region of the corresponding phase space, which is occupied by the reference eddy-resolving solution. 
Based on this concept, we propose to use methods of constrained optimization to confine the low-resolution solution to remain within the correct phase space region, without attempting to amend the eddy physics by introducing a process-based parameterisation.
This approach is advocated as a novel framework for data-driven hyper-parameterisation of mesoscale oceanic eddies in non-eddy-resolving models.
We tested the idea in the context of classical, baroclinic beta-plane turbulence model and showed that non-eddy-resolving solution can be substantially improved towards the reference eddy-resolving benchmark.
\end{abstract}


\end{frontmatter}




\section{Introduction}
The problem of reproducing large-scale flow structures in low-resolution ocean simulations is among the most challenging ones in ocean modelling, and this is mostly due to the lack of information from the small, unresolved scales.
The mainstream approach to this problem is to use parameterisations, that is, mathematically simple and physically justified approximations of the key unresolved and under-resolved small-scale processes (e.g.,~\citet{GentMcwilliams1990,DuanNadiga2007,Frederiksen_et_al2012,JansenHeld2014,
PortaMana_Zanna2014,CooperZanna2015,
Grooms_et_al2015,
Berloff_2016,Berloff_2018,Danilov_etal_2019,Ryzhov_etal_2019,Juricke_etal_2020a,Juricke_etal_2020b,
CCHWS2019_1,Ryzhov_etal_2020,
CCHWS2019_3,CCHWS2020_4,CCHPS2020_J2}).
Despite the decades of research effort invested in this direction, the problem remains mainly unresolved for various reasons, the majority of which is due to inaccurate description of small-scales physics and energy transfers between scales.
 
This work takes a new angle on this long-standing problem.
Namely, we propose to look at the problem from the dynamical system point of view and consider the inability of the low-resolution model to reproduce the large-scale flow structures as the persistent tendency of phase space trajectories representing the low-resolution solution to escape the right region of the corresponding phase space, which is occupied by the reference eddy-resolving solution.
Thus, instead of parameterising directly the action of unresolved scales onto the resolved ones, our approach is to constrain the low-resolution flow dynamics to the right region of the phase space.
Two other methods based on this approach were developed and tested earlier.
The first method computes the low-resolution solution as the phase space trajectory of the image point, which is dynamically governed by the reference solution~\citep{SB2021_J1}.
The second method reconstructs a dynamical system from the  reference solution, and then explicitly uses this system to predict the low-resolutions circulation~\citep{SB2021_J2}.
In this study we propose a new method: the employed low-resolution model, which is integrated explicitly, is constrained by an optimization procedure that restricts the solution to stay within the right region of phase space, as defined from the reference solution.
For the proof-of-concept stage, we consider, first, the Lorenz model, and, then, the classical, baroclinic quasigeostrophic (QG) model of the beta-plane turbulence.

\section{The method\label{sec:lorenz63}}
We run the mathematical model of the studied phenomenon (here, the QG model) at low resolution and restrict its solution to the reference region of phase space defined by a set of constraints $g$ (a ball of radius $r$ in our case).
The method is based on solving a constrained minimization problem of the form (e.g.,~\citet{Bertsekas1996}):
\begin{equation}
 \begin{aligned}
  \min && f(\mathbf{x}),\\
  \text{subject to} && g(\mathbf{x})\le 0 \, ,
 \end{aligned}
 \label{eq:minproblem1}
\end{equation}
where $f$ and $g$ are given functions.
In order to inject the constraint into the minimization problem, we use the barrier function $\phi=-1/g(\mathbf{x})$, which leads to the following unconstrained minimization problem:
\begin{equation}
 \begin{aligned}
  \min && f(\mathbf{x})+\frac{\mu}{g(\mathbf{x})} \,, \quad \mu\rightarrow 0 \, ,\\
 \end{aligned}
 \label{eq:minproblem}
\end{equation}
where parameter $\mu>0$ regulates accuracy for finding the minimum 
In our case, a very accurate approximation of the minimum is not required as the problem is chaotic, and the goal of constraining will be achieved anyway. 
Without compromising the main message, we leave further technicalities beyond the scope of our work and refer readers to~\citet{Bertsekas1996}. 

As an example, we first consider the Lorenz 63 system~\citep{Lorenz1963}:
\begin{equation}
\begin{aligned}
&& \mathbf{x}'(t)=\mathbf{F}(\mathbf{x}(t)),\quad \mathbf{F}:=
\begin{pmatrix}
\sigma(y-x)\\
x(\rho-z)-y\\
xy-\beta z\\
\end{pmatrix},\\
&& \text{subject to}\quad g(\mathbf{x}):=x^2+y^2+z^2-r^2\le0 \, ,\\
\end{aligned}
\label{eq:Lorenz63}
\end{equation}
where prime denotes a derivative with respect to time, and $\mathbf{x}(t)=(x(t),y(t),z(t))$.
We take $\sigma=10$, $\beta=8/3$, $\rho=28$; and the initial condition $\mathbf{x}(t_0)=(-4.32, -6.00, 18.34)$ is chosen close to the Lorenz attractor.

\begin{figure}[h]
\centering
\begin{tabular}{cc}
\begin{minipage}{0.1\textwidth}\hspace*{0.05cm} {\bf (a)} \end{minipage} & \begin{minipage}{0.1\textwidth}\hspace*{0.05cm} {\bf (b)} \end{minipage}\\
\begin{minipage}{0.5\textwidth}\includegraphics[scale=0.21]{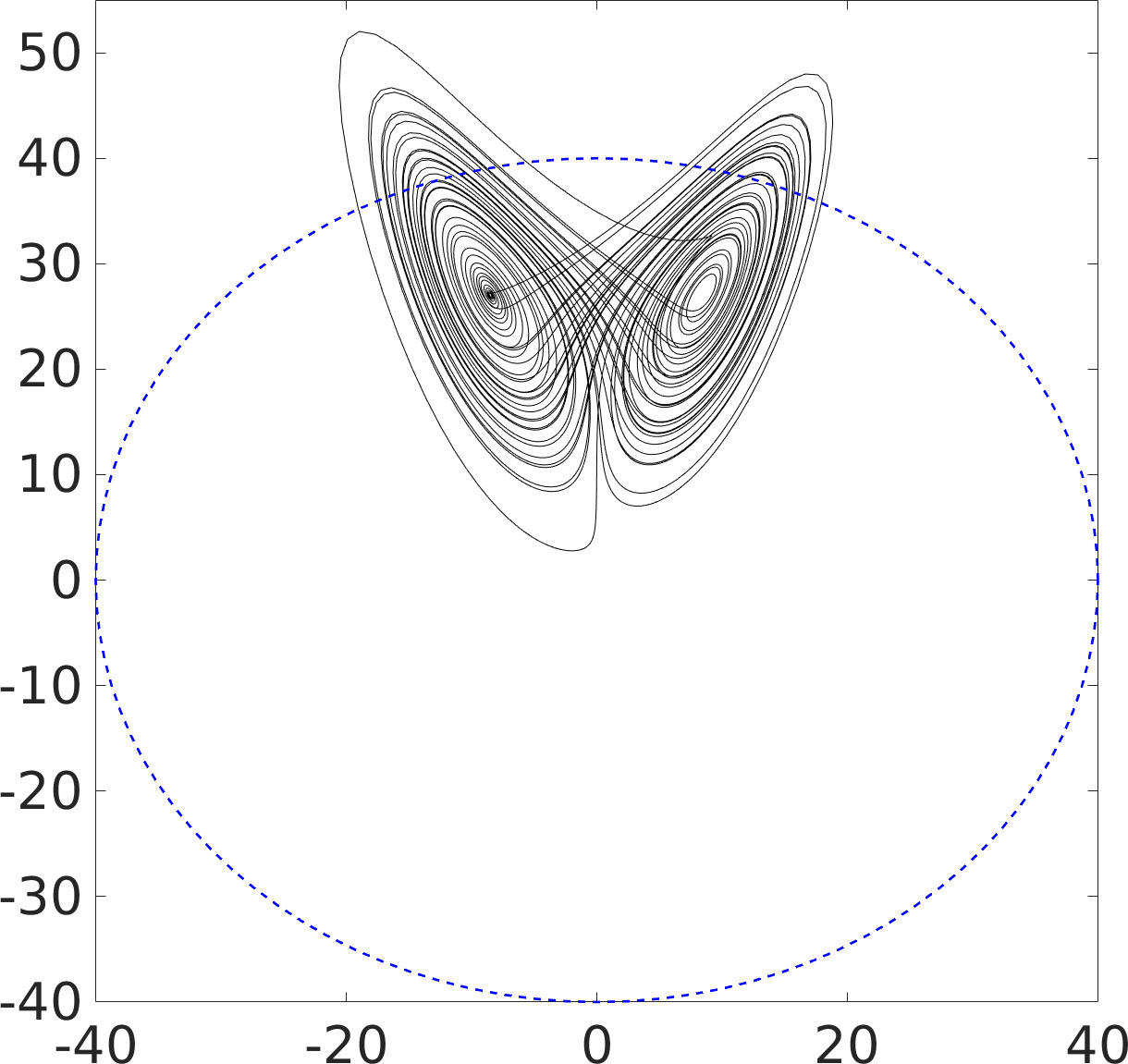}\end{minipage} &
\begin{minipage}{0.5\textwidth}\includegraphics[scale=0.21]{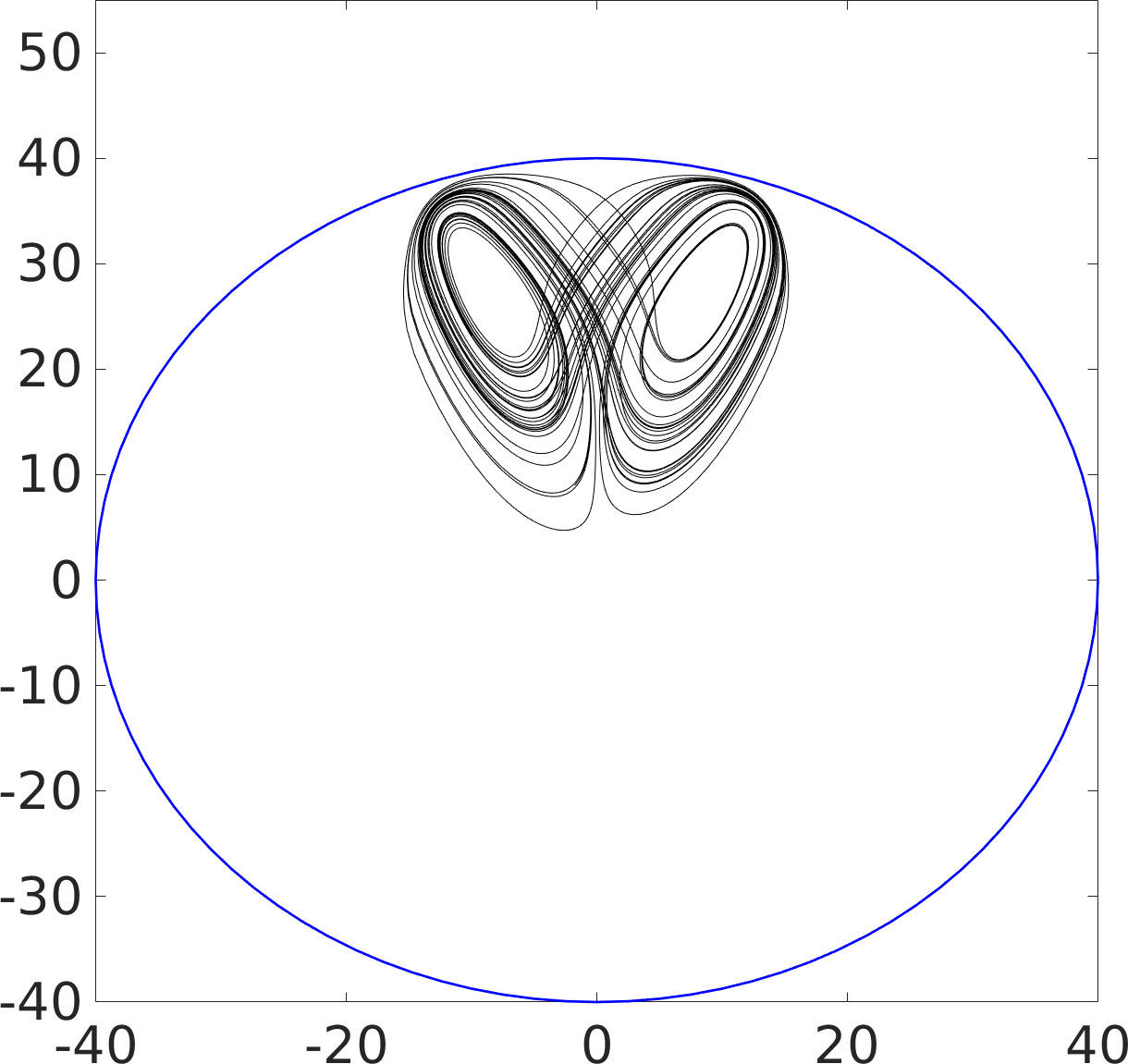}\end{minipage}\\
\end{tabular}
\caption{
Shown are {\bf (a)} solution of the Lorenz system~\eqref{eq:Lorenz63} projected on the (x,z)-plane and {\bf (b)} the corresponding solution constrained to a ball of radius $r=40$ (blue line) centred at $(0,0,0)$ for $t=[0,50]$.
}
\label{fig:Lorenz63}
\end{figure}

As seen in Figure~\ref{fig:Lorenz63}a, the solution to the Lorenz system remains on the attractor over the whole simulation time, and there are no constraints influencing the trajectory.
If trajectory is to be restricted to some region of the phase space, then a set of constraints needs to be introduced.
We chose a ball of radius $r=40$ centred at $(0,0,0)$ to be the constraining phase space region, and the constrained solution remains in it over the whole simulation (Figure~\ref{fig:Lorenz63}b). 
But, the constrained solution is not the same as the original one even within the ball interior -- the region where trajectory 
is allowed to evolve with no constraints as long as the constraints are satisfied.
It happens because the constraints deform the structure of the original phase space 
by imposing extra restrictions on the solution.
If we keep the radius decreasing, the deformation eventually becomes so large that the constrained system completely fails to reproduce the Lorenz attractor (not shown).
This indicates that the Lorenz system should be somehow modified, but discussing this is beyond the scope of our study.

The above example shows how one can use the constrained optimization approach in more sophisticated settings.
For instance, one can take the primitive-equations or a QG ocean model, compute its eddy-resolving reference solution, 
project it onto the coarse grid, and find (approximately) a spherical region of the phase space that is occupied by this solution.
If the unconstrained low-resolution model configuration cannot reproduce the (nominally resolved) reference circulation,
then implementation of constraint is justified, and the questions are whether it works and how can it be tuned and optimized.

\section{Multilayer quasi-geostrophic equations\label{sec:qg}}
In this section we apply the method to the two-layer QG model describing evolution of the potential vorticity (PV) anomaly $\mathbf{q}=(q_1,q_2)$ in a domain $\Omega$~\citep{Pedlosky1987}:
\begin{equation}
\begin{aligned}
\frac{\partial q_1}{\partial t}+\mathbf{u}_1\cdot\nabla q_1&=\nu\nabla^4\psi_1-\beta\frac{\partial \psi_1}{\partial x},\\
\frac{\partial q_2}{\partial t}+\mathbf{u}_2\cdot\nabla q_2&=\nu\nabla^4\psi_2-\mu\nabla^2\psi_2-\beta\frac{\partial \psi_2}{\partial x} \, ,
\end{aligned}
\label{eq:pv}
\end{equation}
where $\boldsymbol{\psi}=(\psi_1,\psi_2)$ is the velocity streamfunction in the top and bottom layers, respectively; $\beta=2\times10^{-11}\, {\rm m^{-1}\, s^{-1}}$ is the planetary vorticity gradient; $\mu=4\times10^{-9}\, {\rm s^{-1}}$ is the bottom friction parameter; $\nu$ is the lateral eddy viscosity (to be specified later), and $\mathbf{u}=(u,v)$ is the flow velocity vector.
The ocean basin $\Omega=[0,L_x]\times[0,L_y]\times[0,H]$ is zonally periodic flat-bottom channel with horizontal dimensions $L_x=1800\, \rm km$ and $L_y=L_x/2$, with the depth $H=H_1+H_2$, and filled out by two stacked isopycnal fluid layers of depths $H_1=1.0\, \rm km$ and $H_2=3.0\, \rm km$.

Forcing in~\eqref{eq:pv} is introduced via a vertically sheared, baroclinically unstable, background flow (e.g.,~\cite{BerloffKamenkovich2013}):
\begin{equation}
\psi_i\rightarrow-U_i\,y+\psi_i,\quad i=1,2 \, , 
\label{eq:forcing}
\end{equation}
with the zonal velocities $U=[6.0,0.0]\,\rm cm\, s^{-1}$.

The layer-wise PV anomalies and streamfunctions are related through the pair of coupled elliptic equations:
\begin{subequations}
\begin{align}
q_1=\nabla^2\psi_1+s_1(\psi_2-\psi_1),\\
q_2=\nabla^2\psi_2+s_2(\psi_1-\psi_2) \, ,
\end{align}
\label{eq:q_psi}
\end{subequations}
with stratification parameters $s_1=4.22\cdot10^{-3}\,\rm km^{-2}$ and $s_2=1.41\cdot10^{-3}\,\rm km^{-2}$ chosen so that the first baroclinic Rossby deformation radius is $Rd_1=25\, {\rm km}$.
The mass and momentum constraints are imposed following McWilliams (1977).
System~(\ref{eq:pv})-(\ref{eq:q_psi}) is augmented by the periodic horizontal boundary conditions set at the eastern ($\Gamma_2$) and western ($\Gamma_4$) boundaries:
\begin{equation}
\boldsymbol{\psi}\Big|_{\Gamma_2}=\boldsymbol{\psi}\Big|_{\Gamma_4}\,,\quad \boldsymbol{\psi}=(\psi_1,\psi_2) \, ,
\label{eq:bc24}
\end{equation}
and no-slip boundary conditions, 
\begin{equation}
\boldsymbol{u}\Big|_{\Gamma_1}=\boldsymbol{u}\Big|_{\Gamma_3}=0 \, ,
\label{eq:bc13}
\end{equation}
are imposed at the northern ($\Gamma_1$) and southern ($\Gamma_3$) boundaries.
We use $513\times257$ uniform spatial grid, take the eddy viscosity value $\nu=25\, {\rm m^2\, s^{-1}}$, and spin up the model from the state of rest to $t=0$ over the time interval $T_{spin}=[-10,0]\, {\rm years}$, so that the statistically equilibrated flow regime is established.
For the low-resolution model, we use the same set-up, except for much coarser grid $129\times65$ and much larger eddy viscosity value $\nu=250\, {\rm m^2\, s^{-1}}$. 

In order to apply the method to the QG equations~\eqref{eq:pv} one has to choose the variable or variables to constrain.
For this role we take the PV anomaly $\mathbf{q}$ that leads to the inequality $g(\mathbf{q})\le0$; however, the method is not limited
to this particular choice, and other constraints can be implemented.
For example, one can constrain the velocity field or individual terms in the equations.

\begin{figure}[H]
\centering
\begin{tabular}{ccccc}
& \hspace*{0.25cm}\begin{minipage}{0.1\textwidth} {\bf (a)} \end{minipage} & 
\hspace*{0.75cm}\begin{minipage}{0.1\textwidth} {\bf (b)} \end{minipage} &
\hspace*{1cm}\begin{minipage}{0.1\textwidth} {\bf (c)} \end{minipage} &
\hspace*{0.75cm}\begin{minipage}{0.1\textwidth} {\bf (d)} \end{minipage}\\
\hspace*{-1.5cm}\begin{minipage}{0.02\textwidth}\rotatebox{90}{$t=5$ years}\end{minipage}  &
\hspace*{-1cm}\begin{minipage}{0.25\textwidth}\includegraphics[scale=0.08]{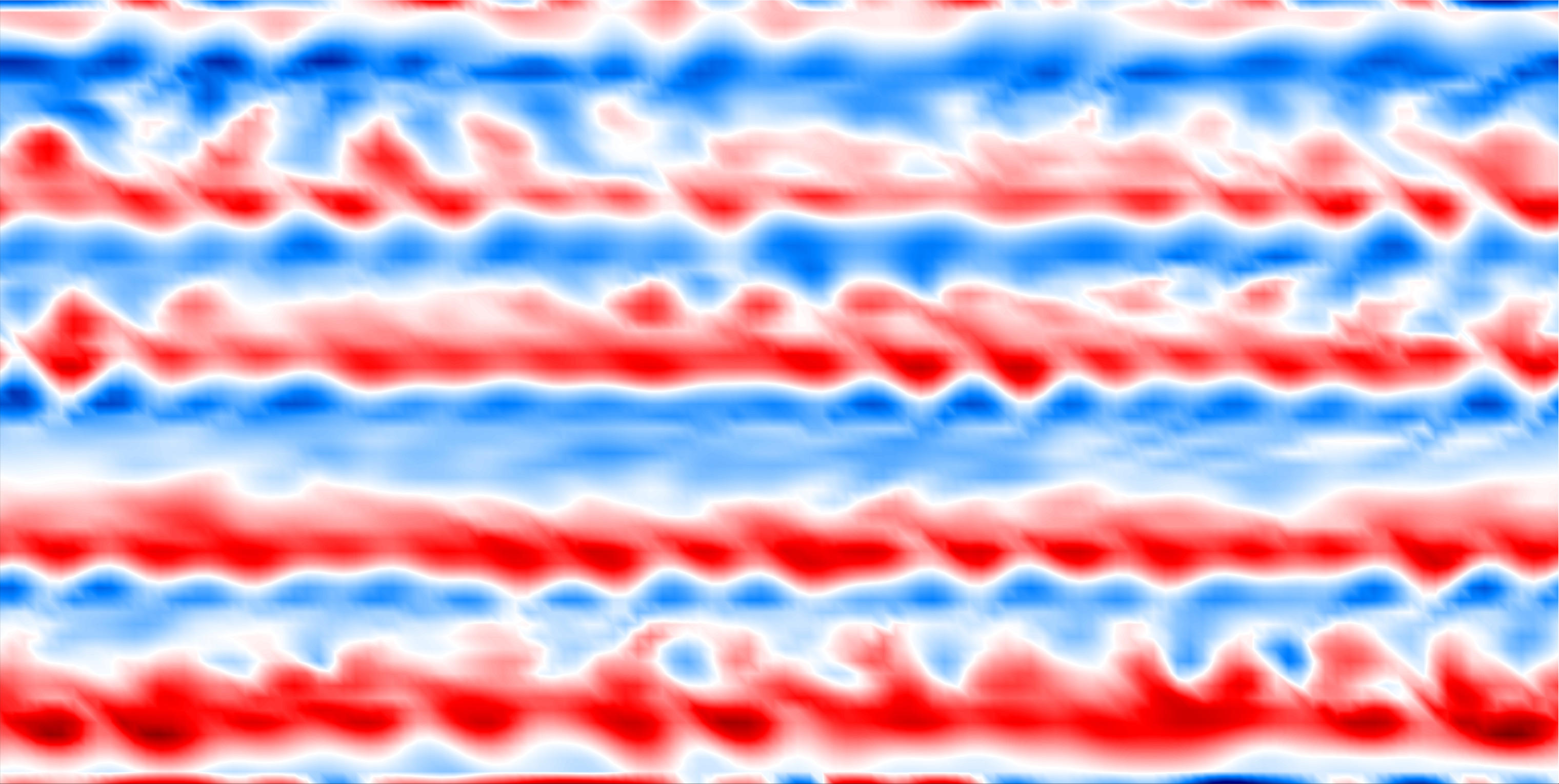}\end{minipage} &
\hspace*{-0.25cm}\begin{minipage}{0.25\textwidth}\includegraphics[scale=0.08]{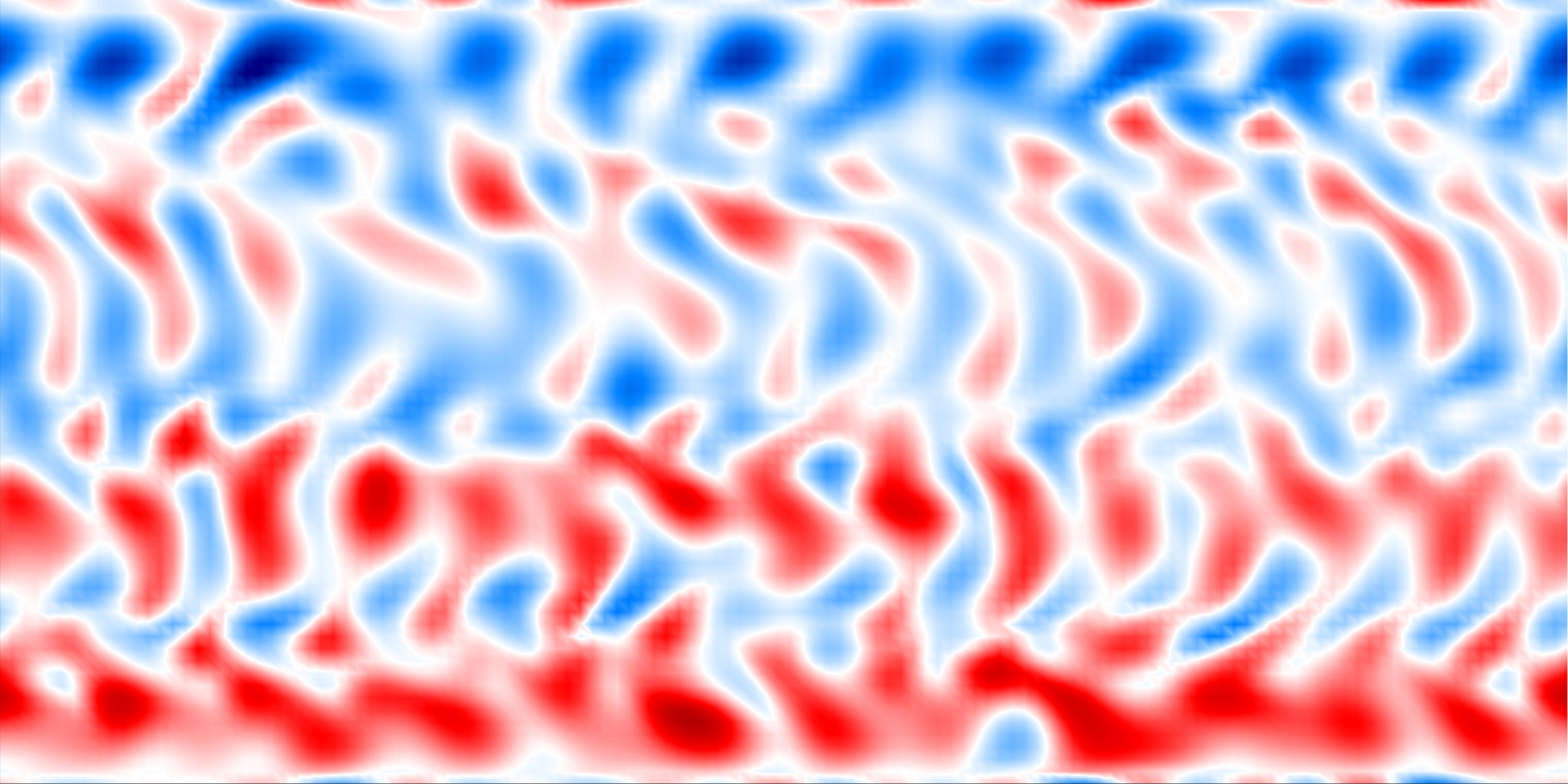}\end{minipage} &
\hspace*{-0.25cm}\begin{minipage}{0.25\textwidth}\includegraphics[scale=0.08]{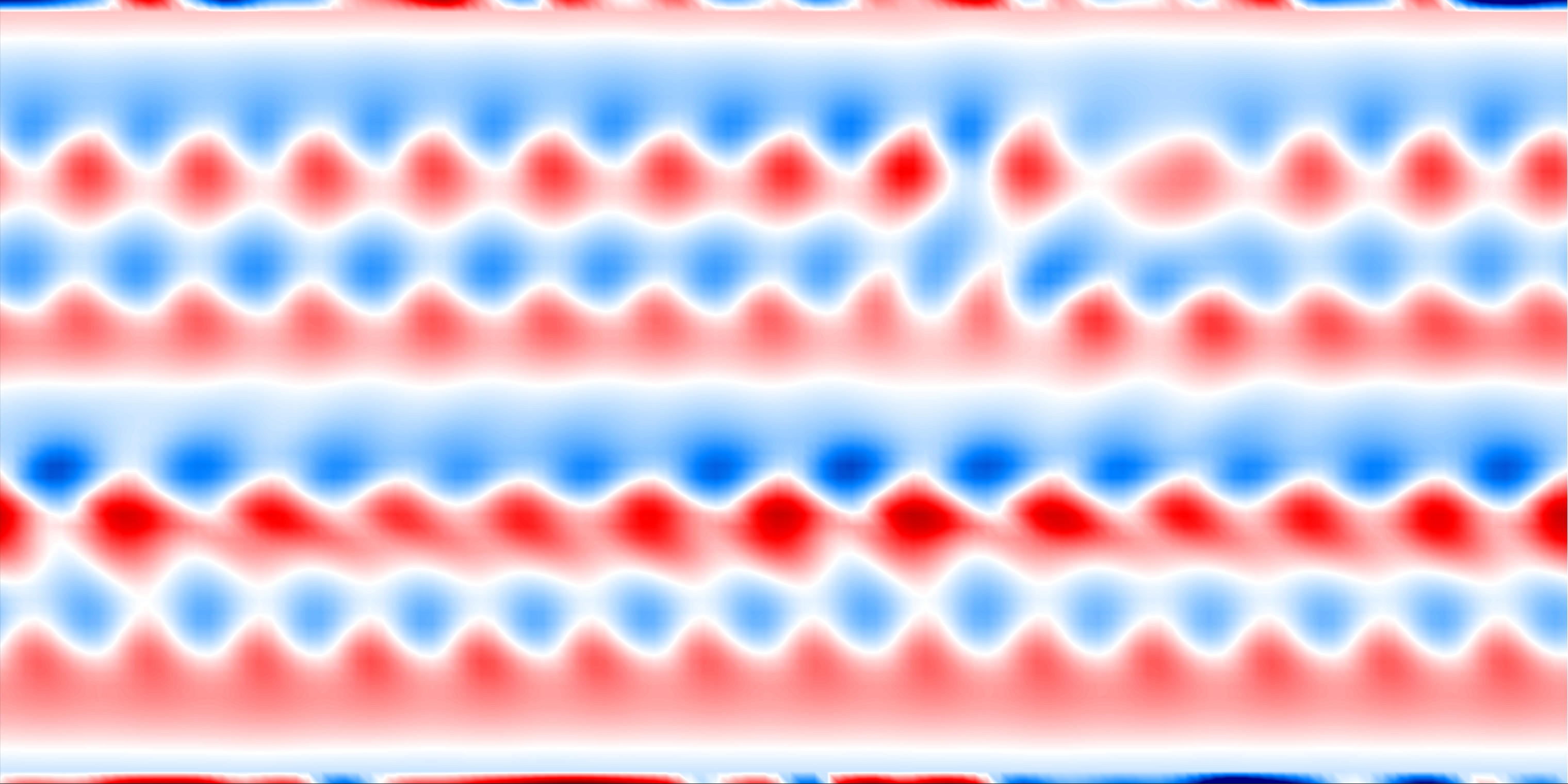}\end{minipage} &
\hspace*{-0.25cm}\begin{minipage}{0.25\textwidth}\includegraphics[scale=0.08]{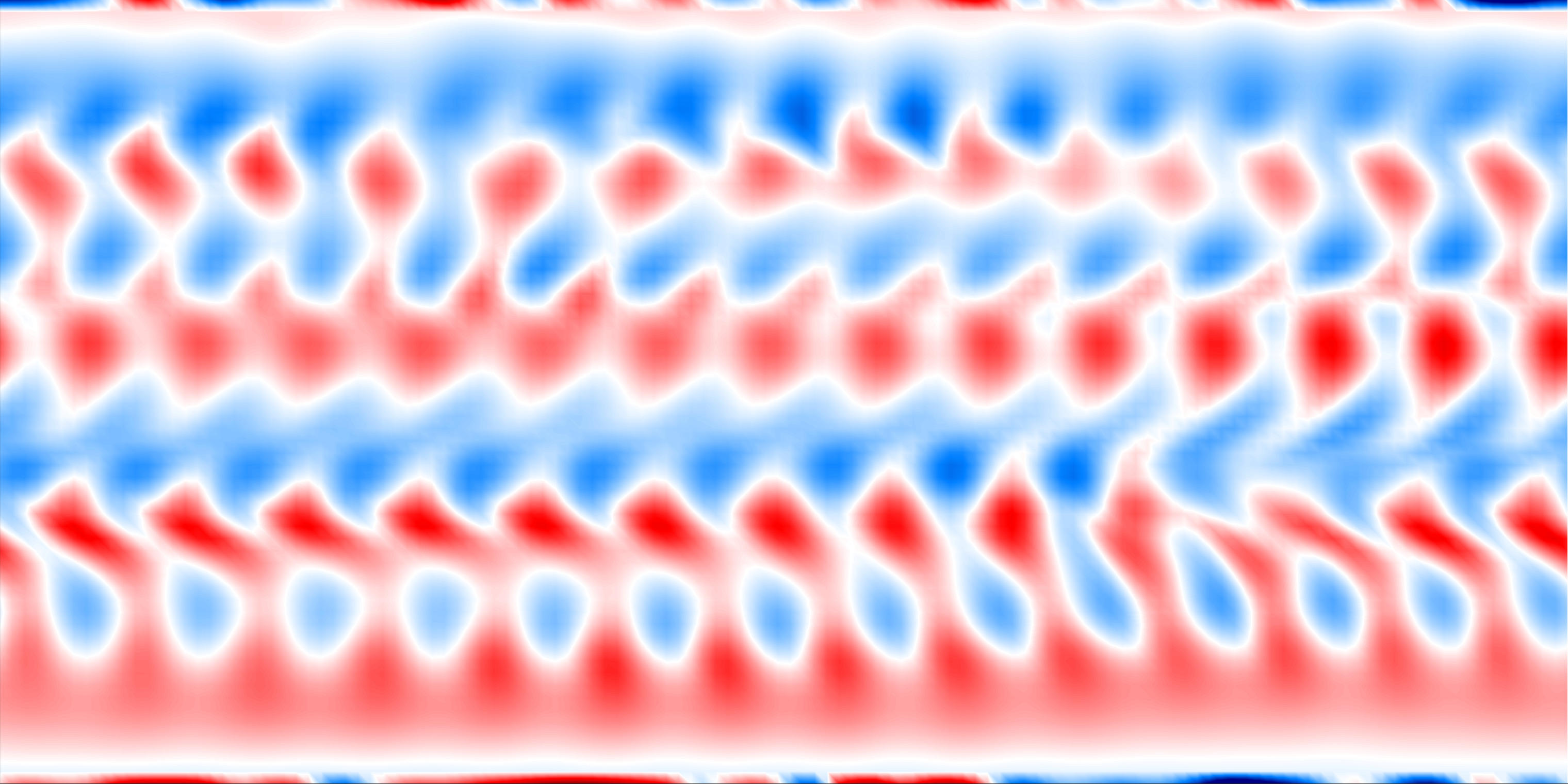}\end{minipage}\\
& & & & \\[-0.35cm]
\hspace*{-1.5cm}\begin{minipage}{0.02\textwidth}\rotatebox{90}{$t=10$ years}\end{minipage}  &
\hspace*{-1cm}\begin{minipage}{0.25\textwidth}\includegraphics[scale=0.08]{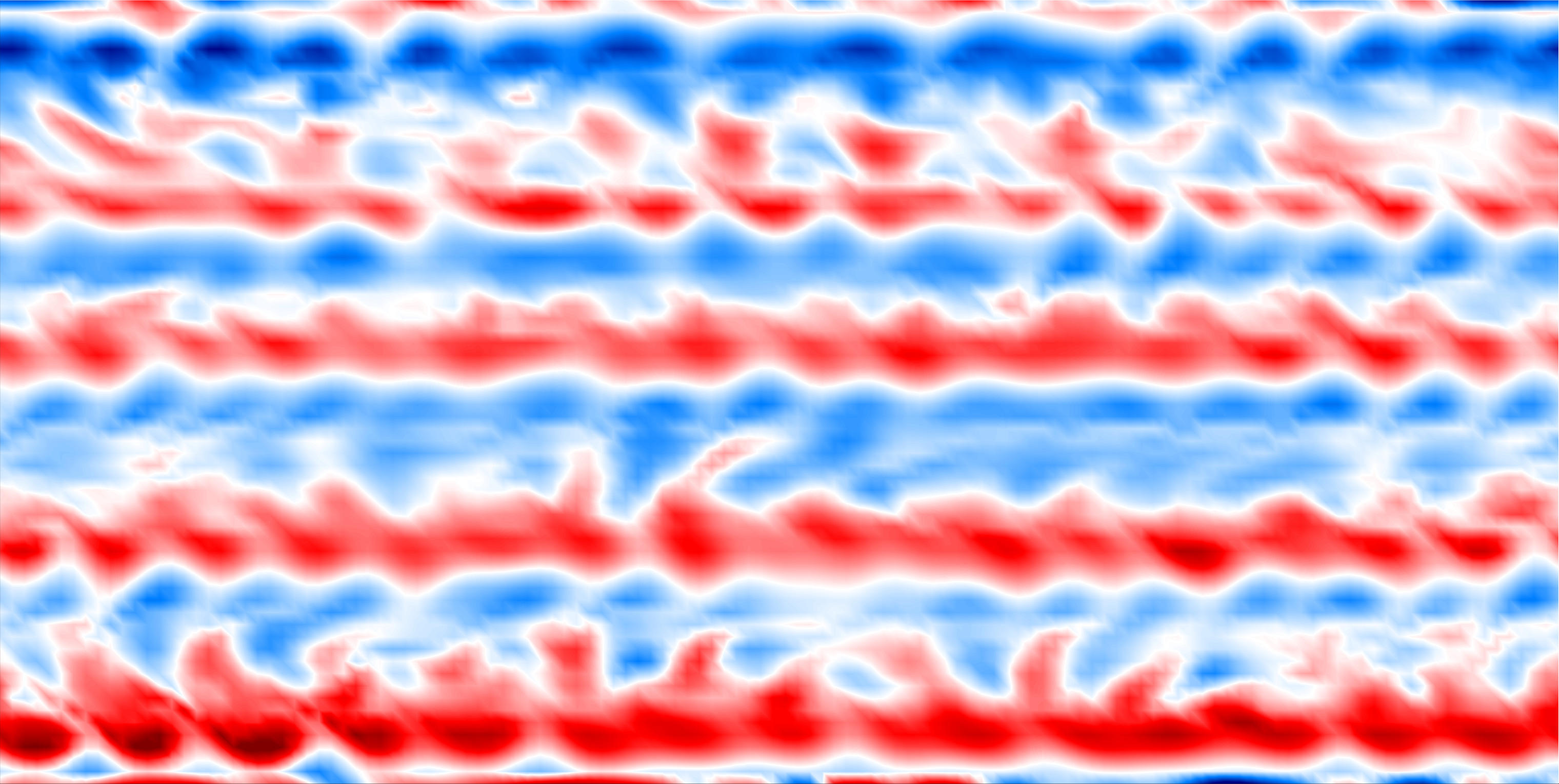}\end{minipage} &
\hspace*{-0.25cm}\begin{minipage}{0.25\textwidth}\includegraphics[scale=0.08]{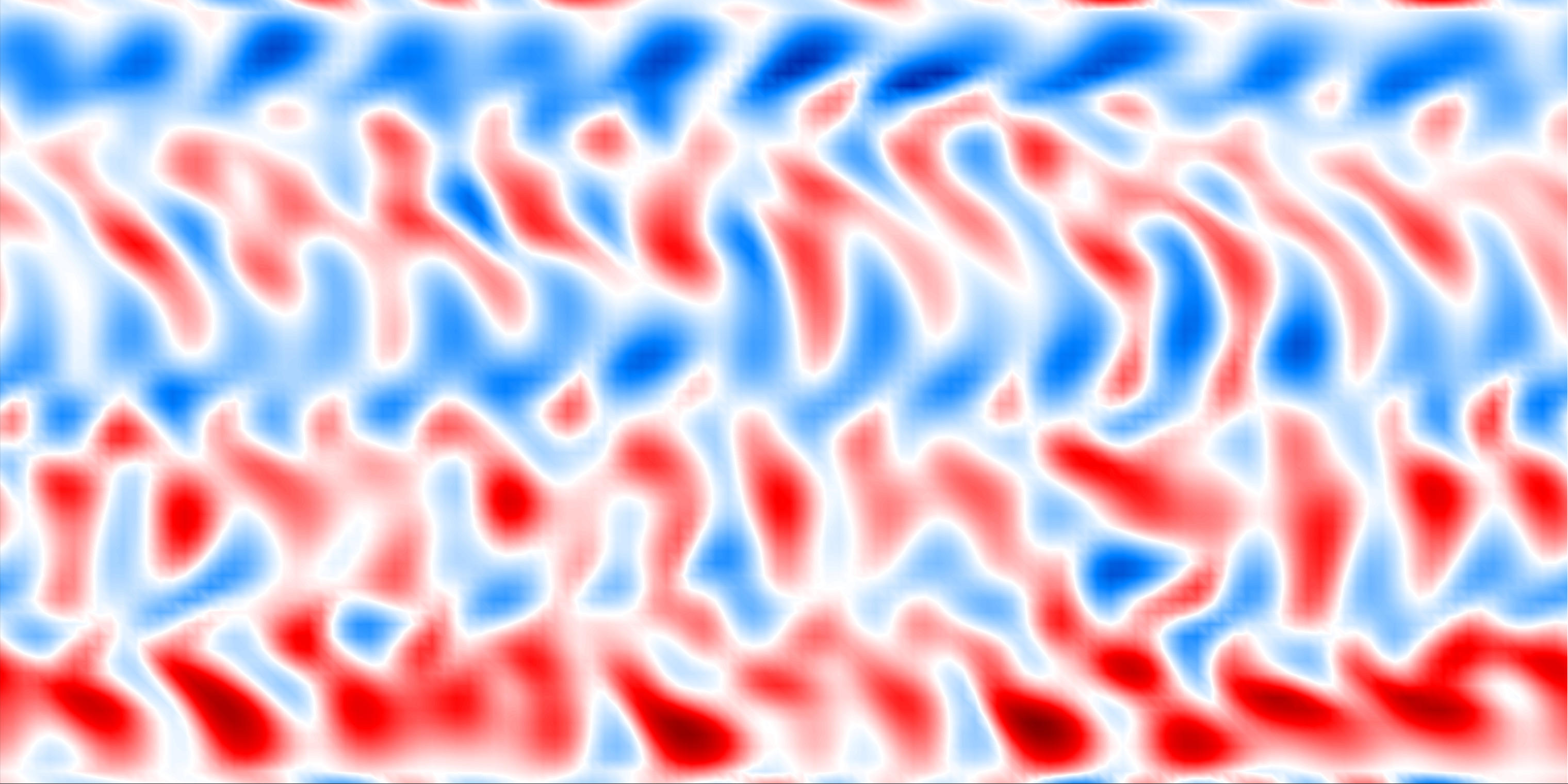}\end{minipage} &
\hspace*{-0.25cm}\begin{minipage}{0.25\textwidth}\includegraphics[scale=0.08]{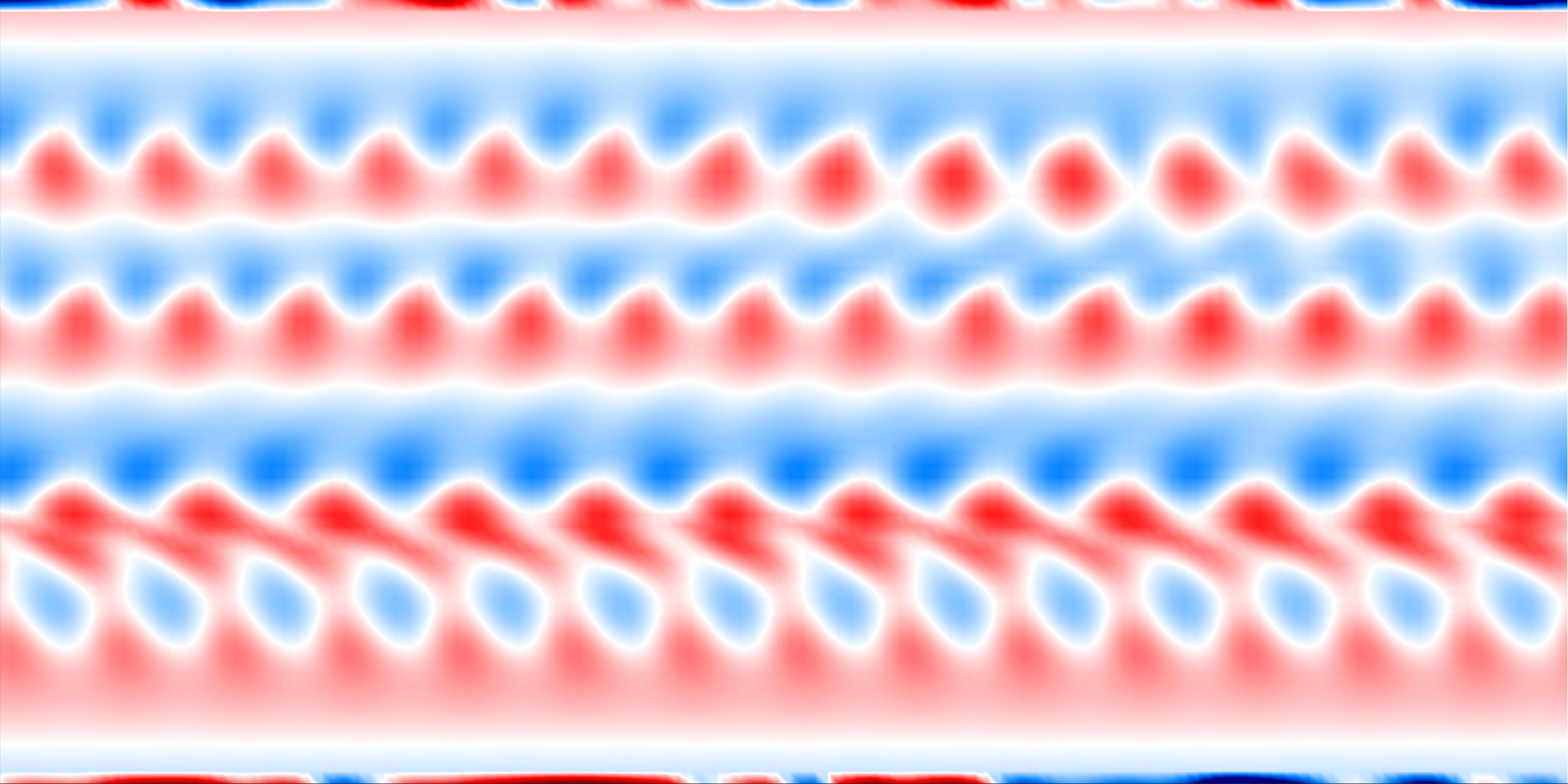}\end{minipage} &
\hspace*{-0.25cm}\begin{minipage}{0.25\textwidth}\includegraphics[scale=0.08]{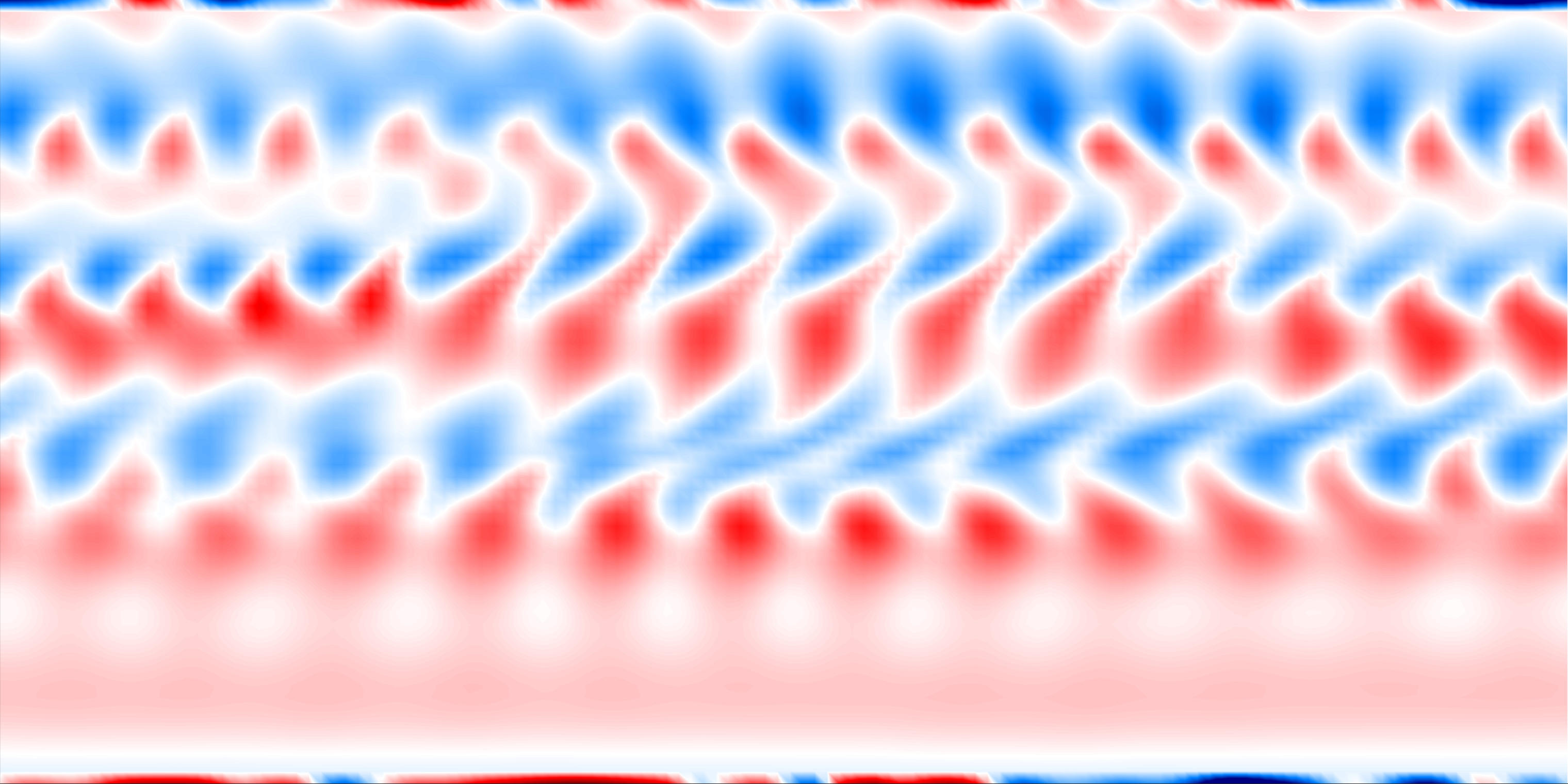}\end{minipage}\\
& & & & \\[-0.35cm]
\hspace*{-1.5cm}\begin{minipage}{0.02\textwidth}\rotatebox{90}{10-year average}\end{minipage}  &
\hspace*{-1cm}\begin{minipage}{0.25\textwidth}\includegraphics[scale=0.08]{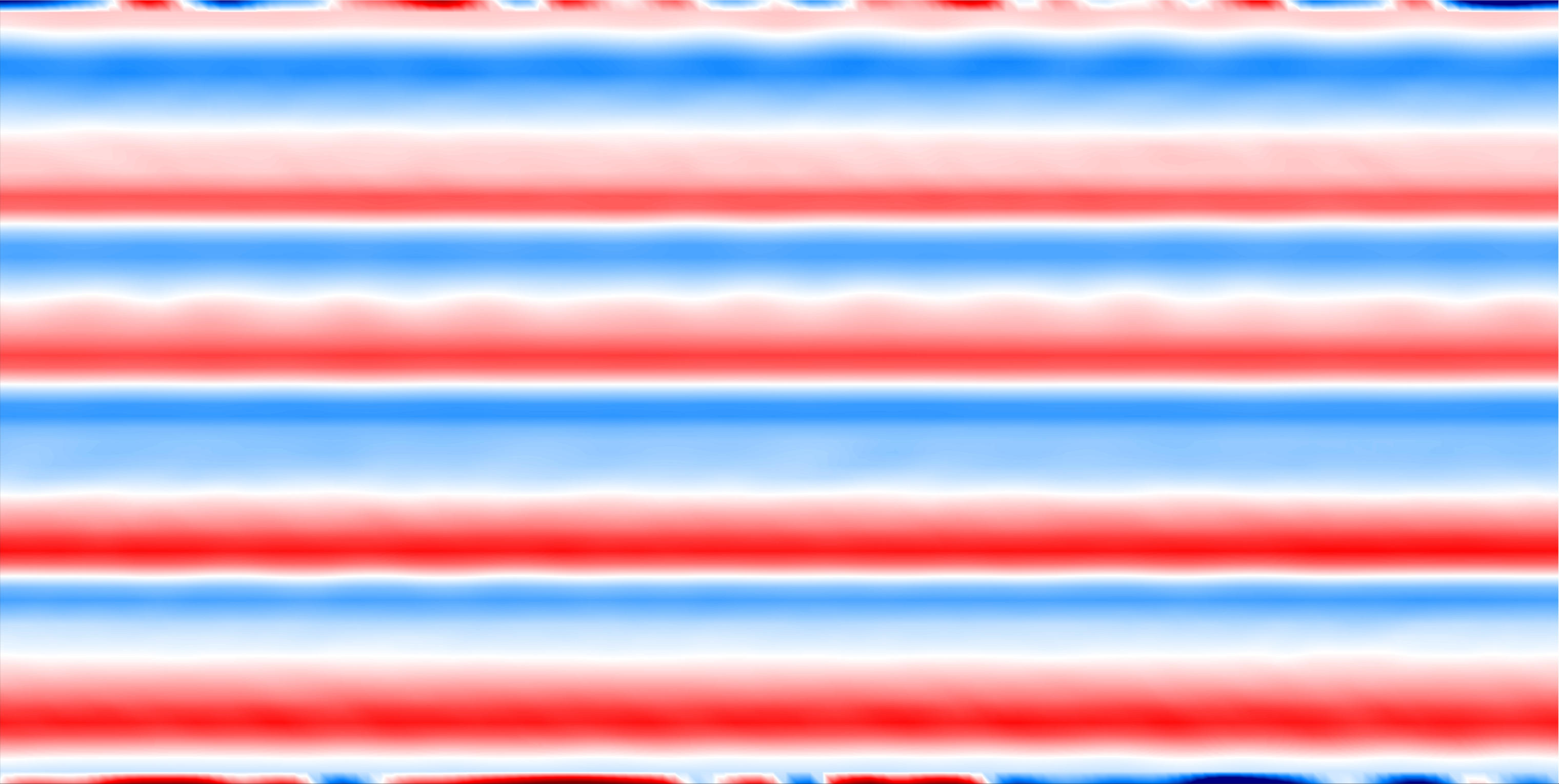}\end{minipage} &
\hspace*{-0.25cm}\begin{minipage}{0.25\textwidth}\includegraphics[scale=0.08]{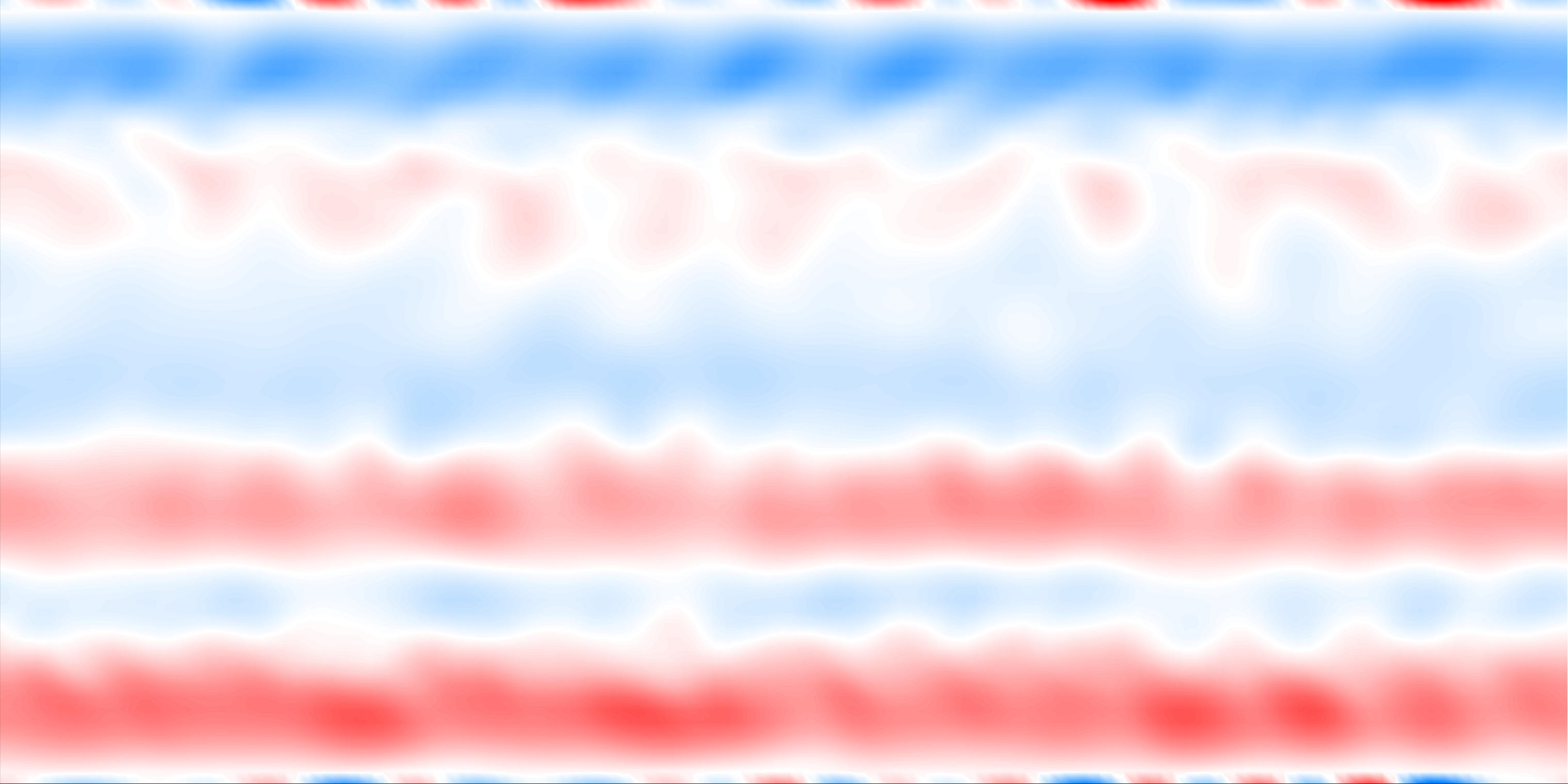}\end{minipage} &
\hspace*{-0.25cm}\begin{minipage}{0.25\textwidth}\includegraphics[scale=0.08]{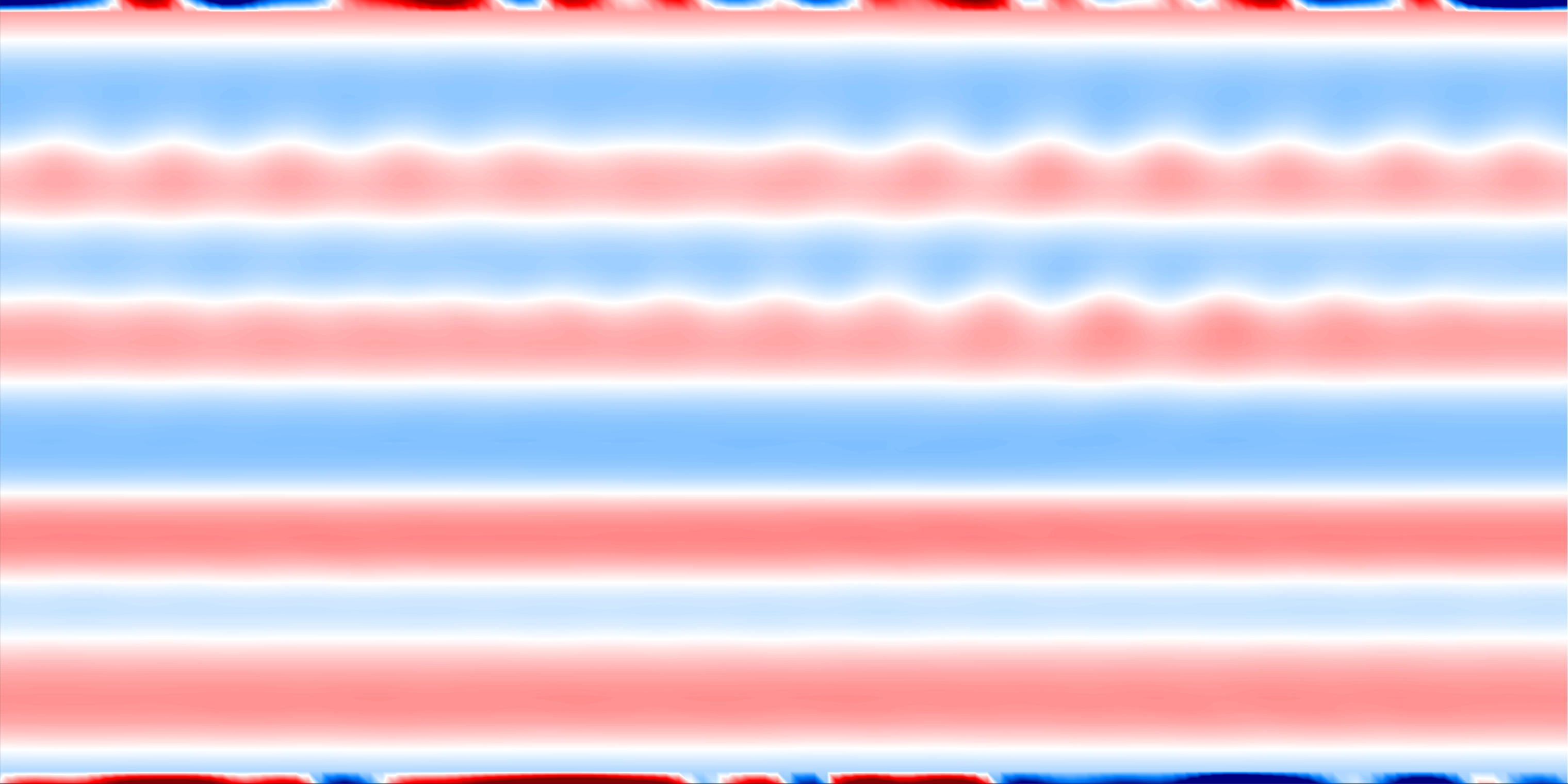}\end{minipage} &
\hspace*{-0.25cm}\begin{minipage}{0.25\textwidth}\includegraphics[scale=0.08]{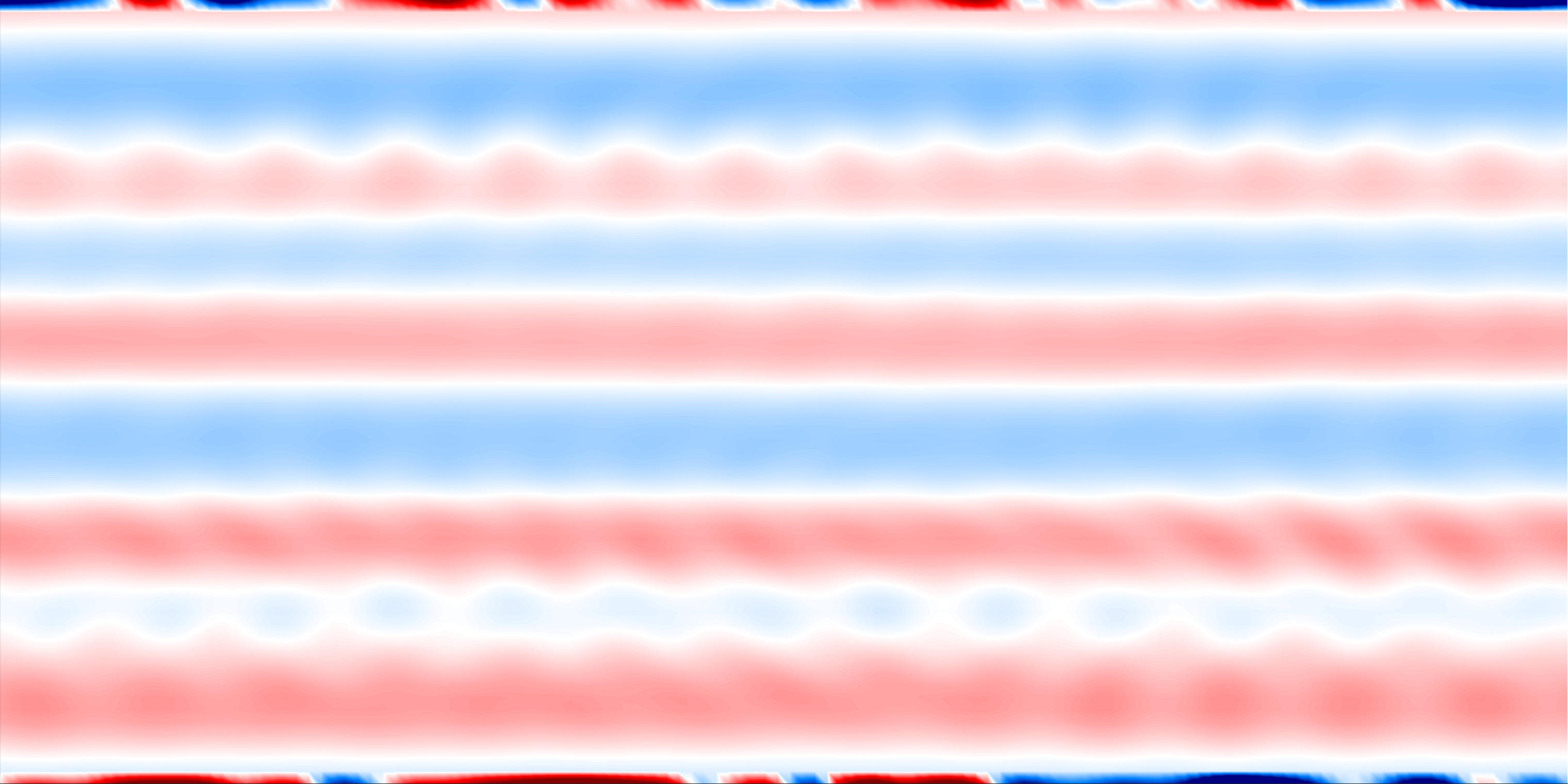}\end{minipage}\\
& & & & \\[-0.5cm]
\multicolumn{5}{c}{\hspace*{-0.5cm}\includegraphics[width=6cm,height=0.75cm]{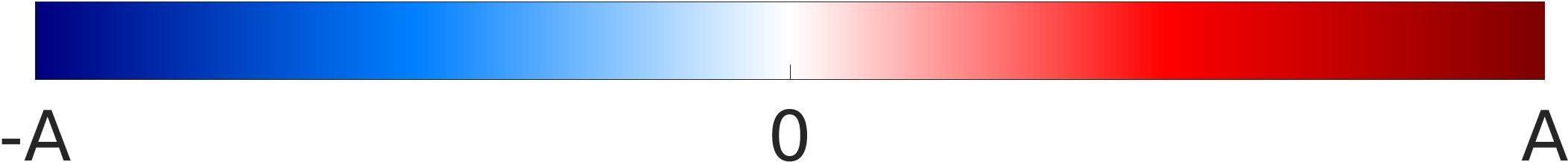}}\\
& & & & \\[-0.35cm]
\hspace*{-1.5cm}\begin{minipage}{0.02\textwidth}\rotatebox{90}{standard deviation}\end{minipage}  &
\hspace*{-1cm}\begin{minipage}{0.25\textwidth}\includegraphics[scale=0.08]{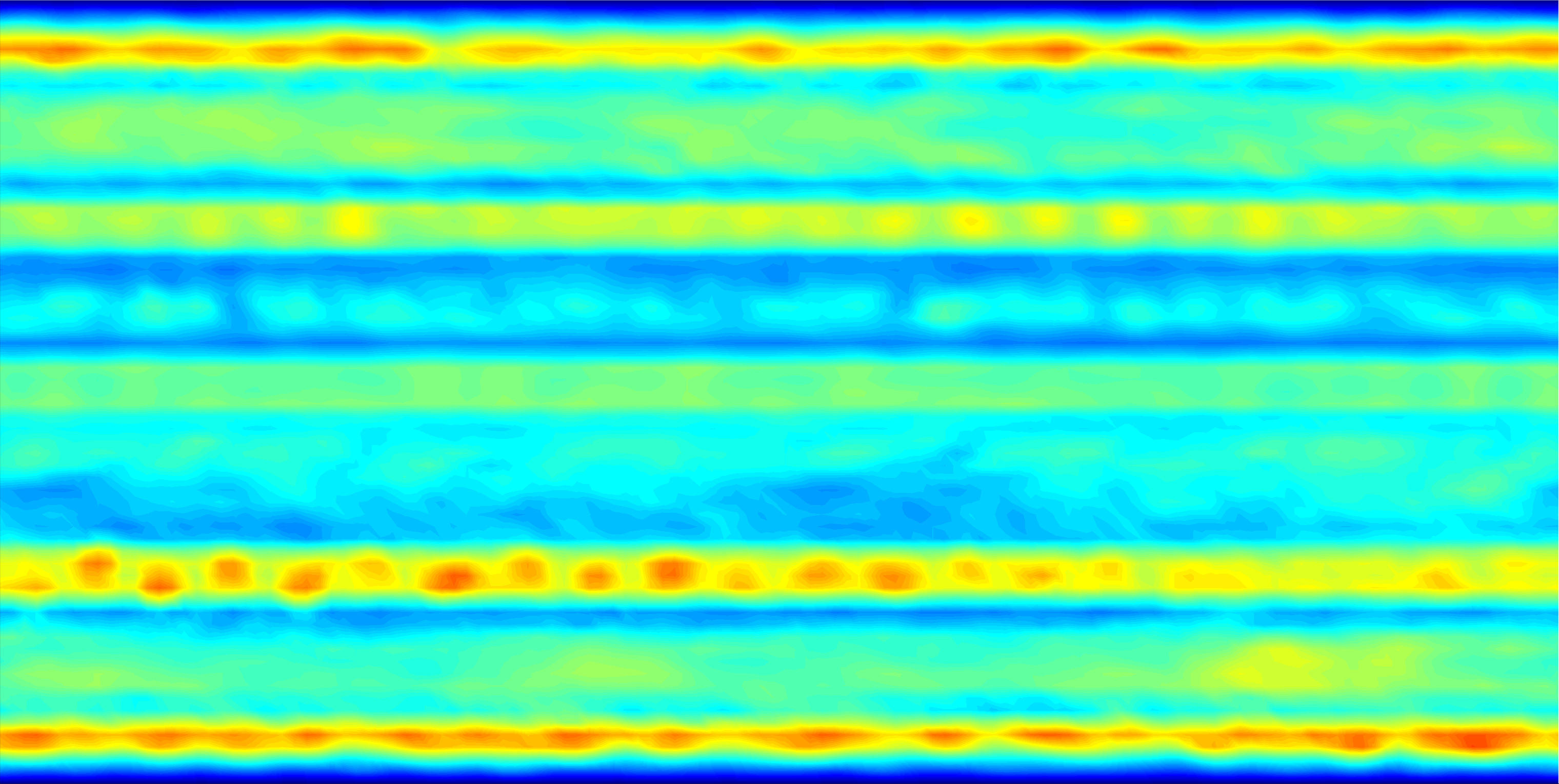}\end{minipage} &
\hspace*{-0.25cm}\begin{minipage}{0.25\textwidth}\includegraphics[scale=0.08]{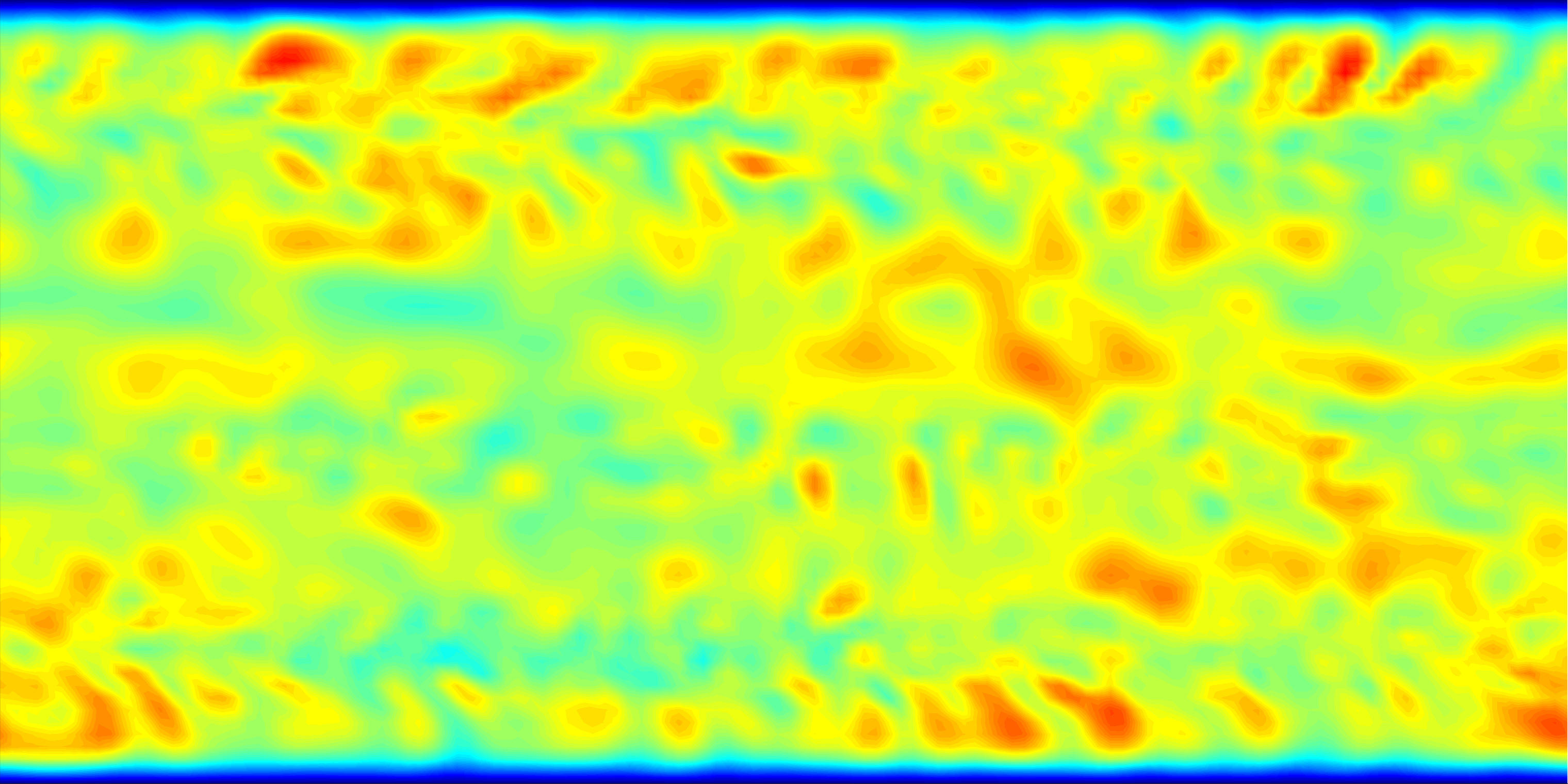}\end{minipage} &
\hspace*{-0.25cm}\begin{minipage}{0.25\textwidth}\includegraphics[scale=0.08]{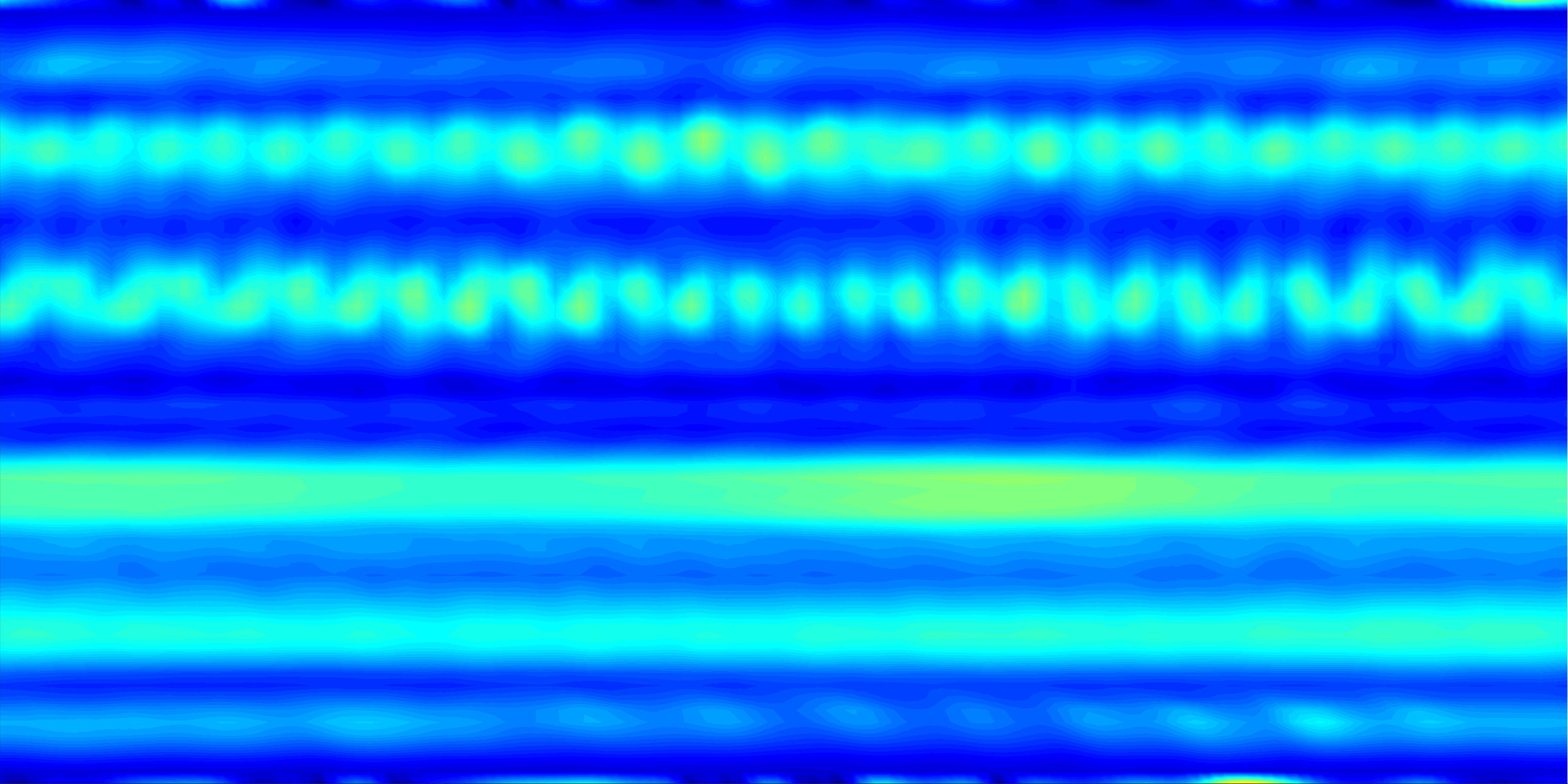}\end{minipage} &
\hspace*{-0.25cm}\begin{minipage}{0.25\textwidth}\includegraphics[scale=0.08]{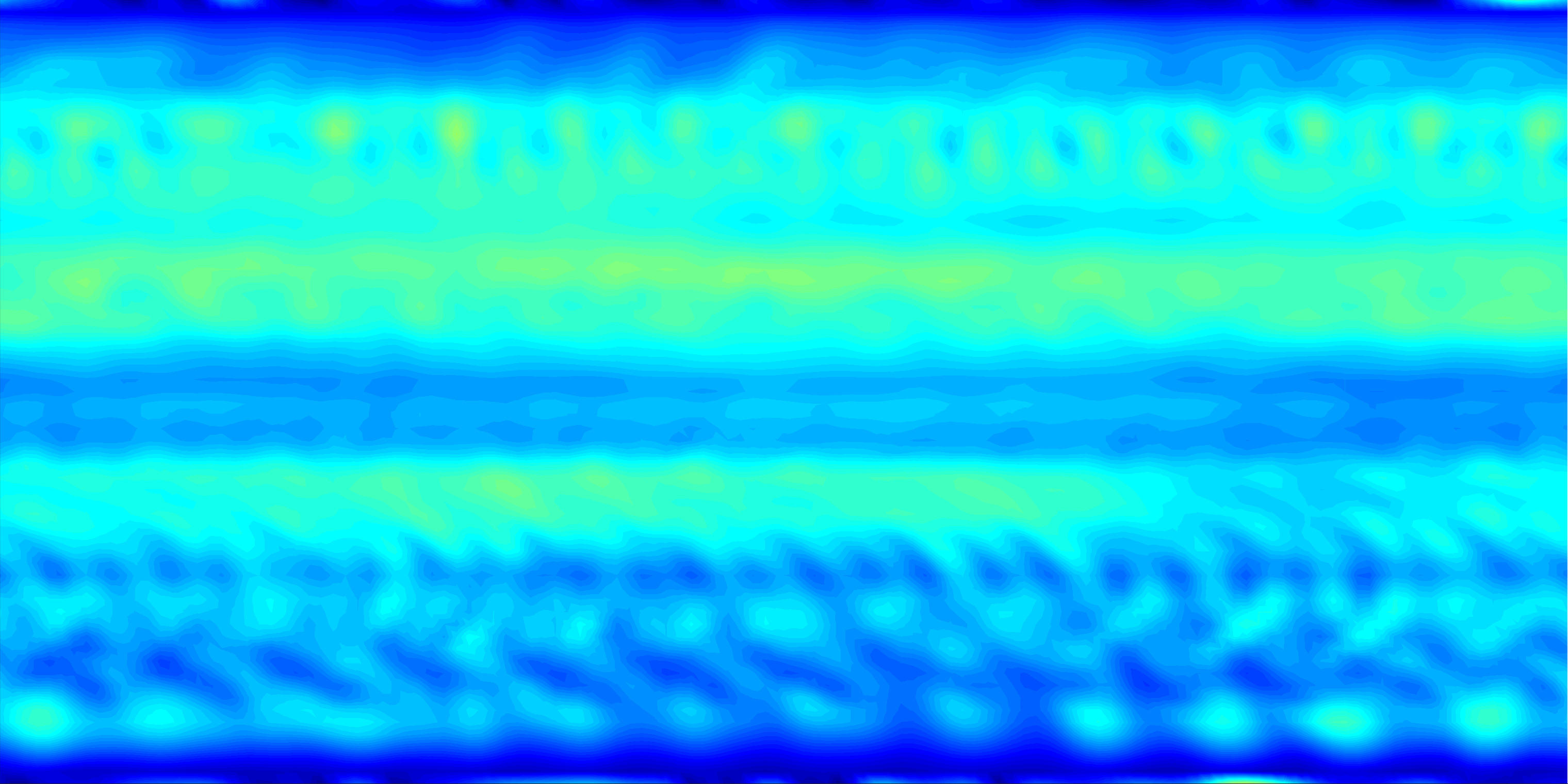}\end{minipage}\\
& & & & \\[-0.75cm]
\multicolumn{5}{c}{\hspace*{-0.5cm}\includegraphics[width=6cm,height=0.75cm]{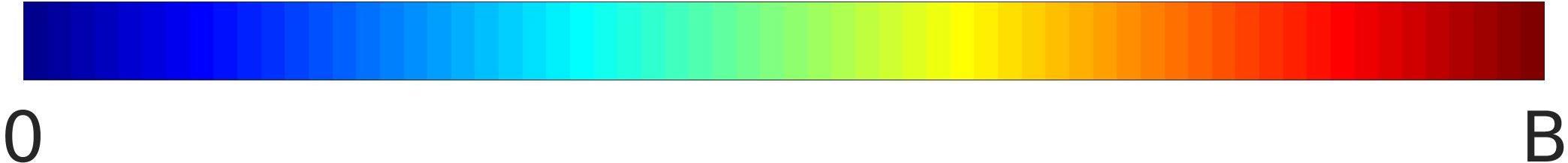}}\\
\end{tabular}
\caption{
Shown are snapshots of the top-layer PV anomaly and standard deviation of:
{\bf (a)} the reference solution (computed with $\nu=25\, {\rm m^2\, s^{-1}}$ on the grid $513\times257$ and then projected on the grid $129\times65$); 
{\bf (b)} the unconstrained low-resolution solution (computed on the coarse grid $129\times65$ with $\nu=250\, {\rm m^2\, s^{-1}}$);
{\bf (c)} the constrained low-resolution solution restricted to the ball with radius $r=846$ (non-dimensional units);
{\bf (d)} the same as {\bf (c)} but for $r=923$.
Units of the PV anomaly fields are non-dimensional; $A=30$, $B=10$.
Note that the constrained solution {\bf(c)} preserves the nominally-resolved flow structures (the 4 jets) both in the snapshots and time mean, whereas the unconstrained one fails on this~{\bf(b)}.
}
\label{fig:qg_sol}
\end{figure}

As seen in Figure~\ref{fig:qg_sol}a, in each direction the reference solution has 4 well-pronounced horizontal jets, which the (unconstrained) low-resolution model fails to reproduce (Figure~\ref{fig:qg_sol}b).
This is because the low-resolution solution trajectory quickly escapes the right phase space region, even when we start it from the right initial condition. 
The phase space is constrained as 
\begin{equation}
g(\mathbf{q}):=(\mathbf{q}-\langle\mathbf{q}\rangle)^2-r^2\le 0,\quad \langle\mathbf{q}\rangle:=\frac1T\int\limits^T_0 \mathbf{q}(t)\, dt,\quad 
r:=\frac1T\int\limits^T_0 \|\mathbf{q}(t)-\langle\mathbf{q}\rangle\|_2\, dt \, ,
\label{eq:qg_const} 
\end{equation}
where $r=846$ (non-dimensional units) and $T=4$ years.
Constraint~\eqref{eq:qg_const} corresponds to the ball of radius $r$ centered at the 4-year time mean of the reference eddy-resolving solution $\langle\mathbf{q}\rangle$; and the radius is found as the mean distance of the 4-year long reference solution from $\langle\mathbf{q}\rangle$.
Here, the ball is the simplest approximation of the real reference-solution attractor in the phase space.
A more accurate approach would be to find an hyper-ellipsoid that takes into account the actual spread of the attractor along different coordinate axes.

When the model is constrained, the 4 jets are recovered both in the instantaneous and time-mean fields (Figure~\ref{fig:qg_sol}c). 
The relative blurriness of the jet edges comes from the use of an order-of-magnitude larger viscosity.
Perhaps, the jet edges can be sharpened up by injecting noise into the advection operator, as in the SALT approach 
(e.g.~\citet{CCHWS2019_1,CCHWS2019_3,CCHWS2020_4,CCHPS2020_J2}), or simply by adding it as a forcing.
To see how the radius of the ball influences the dynamics, we took the ball of a larger radius $r=923$ (this is the maximum radius for the 10-year-long reference solution).
In this case, the constrained solution still has the jets, which are, however, noticeably corrupted (Figure~\ref{fig:qg_sol}d), especially after 10 years of simulation (see the jet near the southern boundary).
Moreover, the separation between the two southern jets becomes less evident (see the time-average subplot) and the standard deviation does not show as pronounced jets as those of the reference solution or the solution constrained in the smaller ball.
This is explained by the fact that the solution constrained by a larger ball drifts farther away from the right phase space region.
Based on this evidence, one could think that a tighter ball would yield a more accurate solution, but this is not necessarily the case, since an over-constrained QG model can lead to significantly incorrect dynamics.
As an example, we found the solution for $r=796$ (this is the minimum radius for the 10-year-long reference solution) and obtained the results very similar to those with $r=846$ (not shown).
This suggests that the ball is likely not an accurate approximation of the reference phase space when finer structures of the solution have to be modelled.
As an alternative, one should focus on methods that can better approximate the reference phase space.

\section{Conclusions and discussion\label{sec:conclusions}}
In this work we have further developed alternative hyper-parameterisation approach for parameterising effects of mesoscale oceanic eddies on the large-scale ocean circulation.
Complimentary to the mainstream physics-based perspective, we propose to deal with the eddy effects from the dynamical systems point of view, and interpret the lack of them as the persistent tendency of phase space trajectories representing the low-resolution solution to escape the right region of the corresponding phase space, which is occupied by the reference eddy-resolving solution. 
Based on this concept, we propose to use methods of constrained optimization to confine the low-resolution solution to remain within the correct phase space region, without attempting to amend the eddy physics by introducing a process-based parameterisation.

We used the Lorenz 63 system as a simple conceptual toy model for the proof of concept. 
Then, we considered the baroclinic, quasigeostrophi (QG) model of the beta-plane turbulence and showed that the low-resolution solution cannot reproduce such nominally-resolved flow structures, as the multiple alternating zonal jets, which are robustly present in the reference eddy-resolving solution.
Implementation of the proposed hyper-parameterisation approach significantly improves the low-resolution model and recovers the jets.
In the reference phase space we used a ball, with the radius found as the time-mean distance of the 4-year-long reference solution from its time mean, and centered at the 4-year time mean of the reference solution.

It is worth drawing reader's attention to the fact that only 4 years of the reference solution have been used to find the centre of the ball and its radius, and this turned out to be enough to reproduce the jets in the 10-year-long run at low resolution.
This demonstrates that the proposed method gives robust results, as soon as it is supplied with sufficient data.
On the other hand, for more sophisticated models or even for different setups of the QG model, longer or shorter runs might be needed to accurately estimate the centre and radius. 

In addition, we have studied how the radius of the ball influences the ability of the constrained QG model to reproduce nominally-resolved structures and found that too large radius eventually lead to the solution degradation, due to the detrimental drift in the phase space, whereas too small radius does not allow for an accurate approximation of the shape of the reference phase space.
Thus, there is the optimal strength of the constraint.

The utility of the proposed method is that it does not require in-depth physical knowledge of interactions between the scales of motion, and it does not require any modification of the governing equations, 
as in the case of traditional parameterisations. 
The method falls into the category of data-driven methods as it requires either detailed observations or their substitute in terms of some eddy-resolving solution data.
Obviously, its advantage may turn into a possible drawback, as often information presented in data is not sufficient (for example, characteristics of the constraining ball may be inaccurately estimated).
This can be mitigated by using more accurate approximations of the constraining geometry (for example, a hyper-ellipsoid instead of the ball to take into account the spread of the solution along different coordinate axes) and of the attractor reconstruction. 
Another intriguing avenue for future research extension is the so called term-wise constraining, which deals with individual dynamical terms, rather than with the whole solution, like in this study.

The proposed method can be implemented into the dynamic core of oceanic general circulation models by constraining solutions of different equations and terms or their combinations.
This is, in turn, the open broad research agenda on the effects of different data-driven constraints 
of the governing equations (e.g.,~\citet{SB2021_J3}).

\section{Acknowledgments}
The authors thank The Leverhulme Trust for the support of this work through the grant RPG-2019-024. 
Pavel Berloff was supported by the NERC grants NE/R011567/1 and NE/T002220/1, and by the Moscow Center for
Fundamental and Applied Mathematics (supported by the Agreement 075-15-2019-1624 with the Ministry of Education
and Science of the Russian Federation).





\bibliographystyle{apalike}
\bibliography{refs}



\end{document}